\documentclass[aps,prb,twocolumn,floatfix]{revtex4-1}
\usepackage{graphicx,txfonts,bm}
\usepackage[colorlinks,citecolor=magenta,linkcolor=blue]{hyperref}

\begin{document}

%% title
%%%%%%%%%%%%%%%%%%%%%%%%%%%%%%%%%%%%%%%%%%%%%%%%%%%%%%%%%%%%%%%%%%%%%%%%%%

\title{Itinerant-localized crossover and orbital dependent correlations for 4$f$ electrons in cerium-based ternary 122 compounds}

%% author list
%%%%%%%%%%%%%%%%%%%%%%%%%%%%%%%%%%%%%%%%%%%%%%%%%%%%%%%%%%%%%%%%%%%%%%%%%%

%% 1st
\author{Haiyan Lu}
\affiliation{Science and Technology on Surface Physics and Chemistry Laboratory, P.O. Box 9-35, Jiangyou 621908, China}

%% 2nd
\author{Li Huang}
\email[Corresponding author: ]{lihuang.dmft@gmail.com}
\affiliation{Science and Technology on Surface Physics and Chemistry Laboratory, P.O. Box 9-35, Jiangyou 621908, China}

%% date information
\date{\today}

%% abstract
%%%%%%%%%%%%%%%%%%%%%%%%%%%%%%%%%%%%%%%%%%%%%%%%%%%%%%%%%%%%%%%%%%%%%%%%%%

\begin{abstract}
The electronic structures of cerium-based ternary 122 compounds Ce$M_{2}$Si$_{2}$, where $M =$ Ru, Rh, Pd, and Ag, are investigated systematically by using the density functional theory in combination with the single-site dynamical mean-field theory. The momentum-resolved spectral functions, total and 4$f$ partial density of states, self-energy functions, and valence state fluctuations are calculated. The obtained results are in good accord with the available experimental data. It is suggested that, upon increasing atomic number from Ru to Ag, the 4$f$ electrons should become increasingly localized. An itinerant-localized crossover for 4$f$ electrons driven by chemical pressure may emerge when $M$ changes from Pd to Ag. Particularly, according to the low-frequency behaviors of 4$f$ self-energy functions, we identify an orbital selective 4$f$ insulating state in CeAg$_{2}$Si$_{2}$, which is totally unexpected.
\end{abstract}

%% make title
%%%%%%%%%%%%%%%%%%%%%%%%%%%%%%%%%%%%%%%%%%%%%%%%%%%%%%%%%%%%%%%%%%%%%%%%%%

\maketitle

%% introduction
%%%%%%%%%%%%%%%%%%%%%%%%%%%%%%%%%%%%%%%%%%%%%%%%%%%%%%%%%%%%%%%%%%%%%%%%%%

\section{introduction\label{sec:intro}}

%% intriguing properties of f electron systems
The cerium-based heavy fermion systems and intermediate valence compounds exhibit a variety of interesting and exotic properties, such as quantum criticality, quantum phase transition, unconventional superconductivity, non-Fermi-liquid behavior, and valence state fluctuation, just to name a few~\cite{RevModPhys.56.755,RevModPhys.63.239}. They have attracted much attention in recent years. It is generally believed that all these features largely originate in the strongly correlated 4$f$ electrons, which manifest Janus-faced behavior (localized or itinerant) depending on surrounding environment. Naturally, an essential question is raised: how and where the 4$f$ electron changes its nature from itinerant to localized, or vice versa? It has been one of the longstanding research issues in the condense matter physics.

%% temperature-driven localized-itinerant crossover
In cerium-based heavy fermion systems and intermediate valence compounds, Kondo temperature $T_{\text{K}}$ is an important energy scale. According to the well-known Doniach phase diagram, cerium's 4$f$ electrons hybridize with conduction electrons and form coherent bands when $T < T_{\text{K}}$. As one would expect, the quasiparticle masses are strongly renormalized. On the contrary, the 4$f$ electrons acquire the localized character and show incoherent electronic fluctuations when $T > T_{\text{K}}$. Clearly, there exists a transition from coherent quasiparticles at low temperature to incoherent fluctuating local moments at high temperature, which is usually called itinerant-localized crossover in the literatures~\cite{PhysRevLett.108.016402,shim:1615,PhysRevB.81.195107}. Very recently, this scenario is directly verified by inelastic neutron scattering measurements and \emph{ab initio} many-body calculations for the dynamic magnetic susceptibility of CePd$_{3}$~\cite{Goremychkin186}.   

%% pressure, magnetic field, chemical pressure driven localized-itinerant crossover
Note that the 4$f$ itinerant-localized crossover is caused not just by temperature changes, but also by pressure, magnetic field, and chemical substitution. For example, it was demonstrated by both experiments and theoretical calculations that pressure can be used to regulate the 4$f$ states of CeIn$_{3}$ from localized to delocalized. As pressure is increased, CeIn$_{3}$ will undergo an electronic Lifshitz transition companied by significant changes in the Fermi surface topology~\cite{PhysRevB.94.075132}. CeRu$_2$Si$_2$ is a typical heavy-fermion compound with a large electronic specific heat coefficient $\gamma$. It takes a metamagnetic transition at $H_m$=7.7~Tesla. Extensive de Haas-van Alphen (dHvA) effect experiments discerned the change of 4$f$ states from itinerant $f$ electrons below $H_m$ to localized $f$ electrons above $H_m$~\cite{JPSJ.83.072001,PhysRevLett.101.056401,PhysRevLett.96.026401,JPSJ.83.072001,JPSJ.61.960,JPSJ.65.515,PhysRevB.90.075127}. Another interesting and well-studied example is the cerium-based ``115'' system, namely Ce$T$In$_{5}$, where $T$ = Co, Rh, and Ir. In CeCoIn$_{5}$ and CeIrIn$_{5}$, the 4$f$ electrons are itinerant. Due to the contributions of coherent 4$f$ electrons, they have enlarged Fermi surfaces. However, in CeRhIn$_{5}$, localized 4$f$ electrons give rise to small Fermi surface. Clearly, chemical composition plays a pivotal role in tuning their 4$f$ states~\cite{PhysRevB.81.195107,shim:1615,PhysRevLett.108.016402}. 

\begin{figure}[t]
\centering
\includegraphics[width=0.45\textwidth]{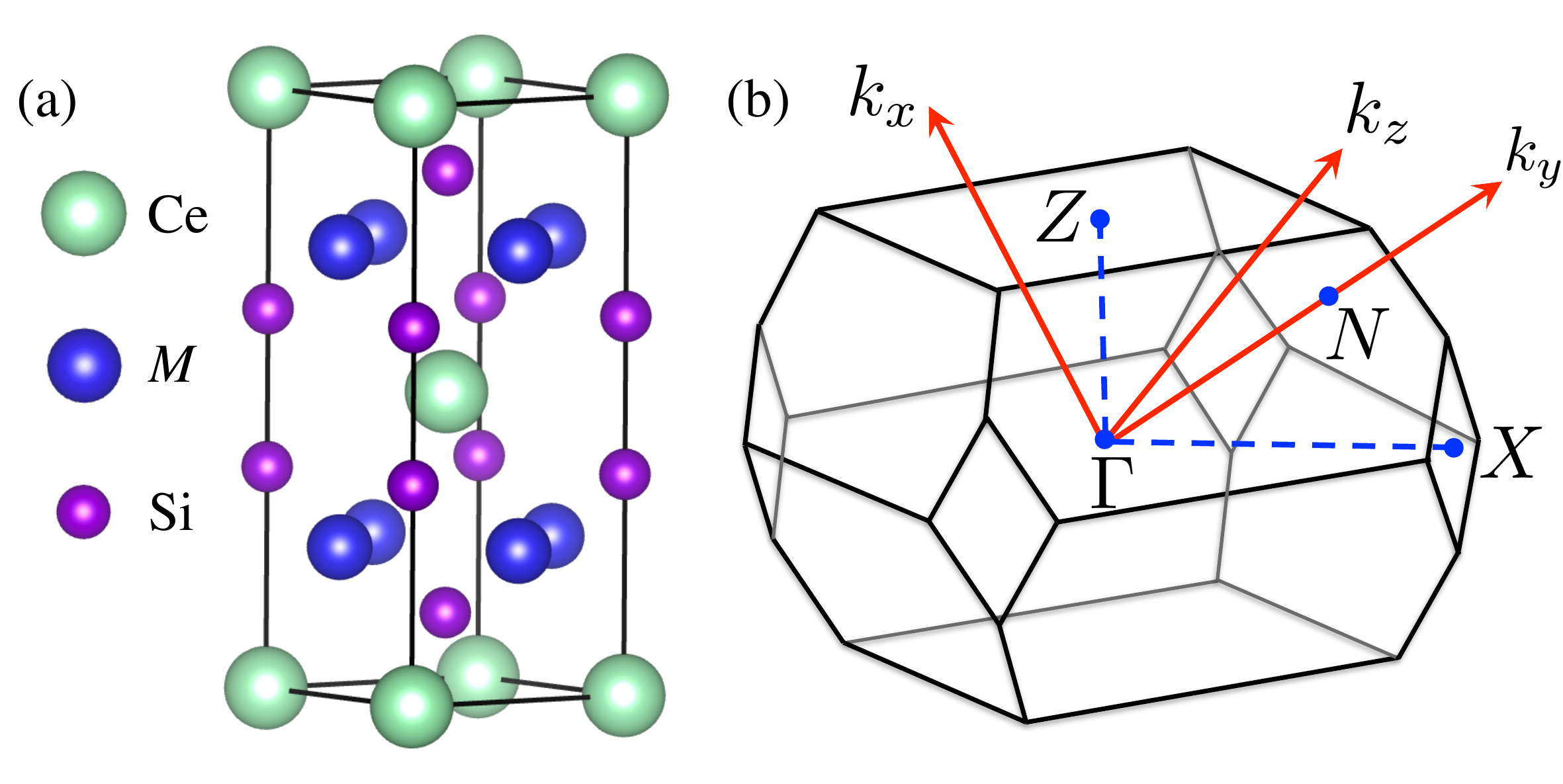}
\caption{(Color online). (a) Schematic crystal structure of Ce$M_{2}$Si$_{2}$ where the Ce, $M$, and Si atoms are represented by green, blue, and purple spheres, respectively. (b) The first Brillouin zone for Ce$M_2$Si$_{2}$ compounds. Some special high-symmetry $k$ points are marked, which will be used in the latter band structure calculations. \label{fig:tstruct}}
\end{figure}

%% our test-bed, Ce-"122" heavy fermion systems, their magnetic properties
In the present study, we would like to concentrate on the cerium-based ternary ``122'' system, namely Ce$M_2$Si$_{2}$, where $M =$ Ru, Rh, Pd, and Ag. We think that these compounds can be considered as valuable supplements to the ``115'' system for examining the chemical substitution driven 4$f$ itinerant-localized crossover. The four compounds crystallize in the body-centered tetragonal ThCr$_2$Si$_{2}$-type structure (see Fig.~\ref{fig:tstruct}), wherein Ce atoms sit on planes well separated by layers of $M$ and Si atoms~\cite{JPSJ.61.2388}. They are notable for the extremely rich magnetic ordered and superconducting phases at low temperature. The ground state of CeRu$_2$Si$_2$ is paramagnetic. As mentioned before, it will undergo a metamagnetic transition at finite magnetic field~\cite{JPSJ.83.072001,PhysRevLett.101.056401,PhysRevLett.96.026401,JPSJ.83.072001,JPSJ.61.960,JPSJ.65.515,PhysRevB.90.075127}. CeRh$_2$Si$_2$ develops complicated antiferromagnetic order below 36~K, which is the highest N\'{e}el temperature among cerium-based heavy-fermion compounds~\cite{PhysRevB.58.8634}. Furthermore, it would turn into superconductor with $T_{c} \sim$ 350~mK~\cite{PhysRevB.53.8241} at pressure above 9~kbar. Likewise, CePd$_2$Si$_2$ transforms into antiferromagnetic state with a staggered magnetic moment below 10~K~\cite{PhysRevB.29.2664} and displays a pressure-induced superconductivity in the range 2 $\sim$ 7~GPa~\cite{PhysRevB.61.8679}. CeAg$_2$Si$_2$ also exhibits antiferromagnetic ground state with $T_{\text{N}} = 8.6$~K. This magnetic ordered phase is completely suppressed when $p \sim 13$~GPa and superconductivity emerges when $p \sim 11$~GPa. The maximal $T_c$ is 1.25~K at $p = 16$~GPa~\cite{SCHEERER2018150}.

%% review the literatures
%% first, CeRu2Si2
%% second, the other compounds
In this cerium-based ``122'' system, undoubtedly, CeRu$_2$Si$_2$ has attracted most of attentions. Exhaustive experiments (including thermodynamic, transport, and spectroscopic measurements)~\cite{PhysRevLett.101.056401, JPSJ.83.072001, JPSJ.65.515, JPSJ.83.072001, PhysRevB.57.R11054, PhysRevLett.96.026401, PhysRevB.40.11429, PhysRevB.90.075127, PhysRevB.85.035127, PhysRevLett.58.820,PhysRevB.42.4329, PhysRevB.56.13689, PhysRevB.86.125138, PhysRevB.44.814, JPSJ.61.960} and theoretical calculations~\cite{JPSJ.62.592, PhysRevB.65.035114, JPSJ.79.024705} have been conducted to unveil the evolution of its 4$f$ states before and after the metamagnetic transition. It is widely accepted that the metamagnetic transition is accompanied by a 4$f$ itinerant-localized crossover. Under small magnetic field, 4$f$ states indeed contribute to the construction of the Fermi surface, manifesting the itinerant 4$f$ electrons. This was confirmed recently by angle-resolved photoemission spectroscopy (ARPES) experiments and band structure calculations for the Fermi surfaces of CeRu$_2$Si$_2$~\cite{PhysRevB.77.035118, PhysRevLett.102.216401}. However, neutron diffraction experiments and static magnetization measurements~\cite{PhysRevB.51.12030} don't support this picture. The results indicated that the itinerant character of the 4$f$ electrons remains almost unchanged during the metamagnetic transition. Hence, this issue is still controversial until now. As for the other compounds in this series, we know a little about their 4$f$ states. Actually, concerning their electronic structures, experimental results are rarely reported in the literatures~\cite{PhysRevB.27.6052}. On the theoretical side, Vildosola \emph{et al.} studied the spectral properties of Ce$M_2$Si$_{2}$ (where $M$=Ru, Rh, and Pd) compounds by means of local density approximation (LDA) combined with Anderson impurity model (AIM). The model was solved within extended non-crossing approximation (NCA)~\cite{PhysRevB.69.125116,PhysRevB.71.184420}. The magnetic quantum critical point of Ce$M_2$Si$_{2}$ was also interpreted by M. Matsumoto \emph{et al.} by solving the Kondo lattice model with the dynamical mean-field theory (DMFT)~\cite{PhysRevLett.103.096403}. In these calculations, the physical models were somewhat oversimplified. Furthermore, the many-body electronic correlation among 4$f$ electrons, spin-orbit coupling, and crystal-field splitting had not been fully taken into accounts. Therefore, it was impossible to obtain reliable results for the detailed electronic structures and related physical properties of Ce$M_2$Si$_{2}$. In this regard, a comprehensive study of the electronic structures of Ce$M_2$Si$_{2}$ by \emph{ab initio} calculations is highly desirable. 

%% the purpose of this paper
In the present paper, we endeavor to uncover the electronic structures of Ce$M_2$Si$_{2}$ by employing a first-principles many-body approach, namely the density functional theory in combination with the single-site dynamical mean-field theory (dubbed as DFT + DMFT)~\cite{RevModPhys.78.865, RevModPhys.68.13}. The band structures, density of states, self-energy functions, and 4$f$ electronic configurations of Ce$M_2$Si$_{2}$ are calculated. We successfully reproduce the ARPES spectra of CeRu$_2$Si$_{2}$. The other results can be viewed as critical predictions. We find that the 4$f$ localized character increases when $M$ goes from Ru to Ag. A chemical pressure driven 4$f$ itinerant-localized crossover is observed when $M$ changes from Pd to Ag. Especially, there exists an orbital selective 4$f$ insulating state in CeAg$_{2}$Si$_{2}$. These results will greatly enrich our knowledge about the 4$f$ states in cerium-based strongly correlated materials. 

%% structure of this paper
The rest of this paper is organized as follows. In Sec.~\ref{sec:method}, the DFT + DMFT computational details are introduced. In Sec.~\ref{sec:results}, the calculated results, including the electronic band structures, total and partial 4$f$ density of states, self-energy functions, and 4$f$ valence state fluctuations are presented and discussed. A detailed comparison between the calculated results and the available experimental data is also provided in this section. Finally, Sec.~\ref{sec:summary} serves as a brief conclusion.  

%% method
%%%%%%%%%%%%%%%%%%%%%%%%%%%%%%%%%%%%%%%%%%%%%%%%%%%%%%%%%%%%%%%%%%%%%%%%%%

\section{method\label{sec:method}}

%% overview of DFT + DMFT
The DFT + DMFT method combines realistic band structure calculation by DFT with non-perturbative many-body treatment of local interaction effects in DMFT~\cite{RevModPhys.68.13,RevModPhys.78.865}. It has been successfully applied to investigate the physical properties of many cerium-based heavy fermion materials in recent years~\cite{Goremychkin186,PhysRevLett.112.106407,shim:1615,PhysRevLett.108.016402,PhysRevB.95.155140,PhysRevB.94.075132,PhysRevB.81.195107}. Here we adopted the DFT + DMFT method to perform charge fully self-consistent calculations to explore the detailed electronic structures of Ce$M_2$Si$_{2}$. The self-consistent implementation of this method is divided into DFT and DMFT parts, which are solved separately by using the \texttt{WIEN2K} code~\cite{wien2k} and the \texttt{EDMFTF} package~\cite{PhysRevB.81.195107}. 

%% computational details
In the DFT part, the experimental crystal structures were used~\cite{STEEMAN1988103}. The generalized gradient approximation was adopted to formulate the exchange-correlation functional~\cite{PhysRevLett.77.3865}. The spin-orbit coupling was taken into account in a second-order variational manner. The $k$-points mesh was $14 \times 14 \times 14$ and $R_{\text{MT}}K_{\text{MAX}} = 7.0$. In the DMFT part, cerium's 4$f$ orbitals were treated as correlated. The four-fermions interaction matrix was parameterized using the Coulomb interaction $U = 6.0$~eV and the Hund's exchange $J_H=0.7$~eV~\cite{PhysRevB.92.104420} via the Slater integrals~\cite{PhysRevB.59.9903}. The fully localized limit scheme was used to calculate the double-counting term for impurity self-energy function~\cite{jpcm:1997}. The constructed multi-orbital Anderson impurity models were solved using the hybridization expansion continuous-time quantum Monte Carlo impurity solver (dubbed as CT-HYB)~\cite{RevModPhys.83.349,PhysRevLett.97.076405,PhysRevB.75.155113}. Note that we not only utilized the good quantum numbers $N$ (total occupancy) and $J$ (total angular momentum) to classify the atomic eigenstates, but also made a severe truncation ($N \in$ [0,3]) for the local Hilbert space~\cite{PhysRevB.75.155113} to reduce the computational burden. Since the inverse temperature $\beta = 100$ ($T \sim 116.0$~K), it was reasonable to retain only the paramagnetic solutions. The convergence criteria for charge and energy were $10^{-4}$ e and $10^{-4}$ Ry, respectively. 

%% results
%%%%%%%%%%%%%%%%%%%%%%%%%%%%%%%%%%%%%%%%%%%%%%%%%%%%%%%%%%%%%%%%%%%%%%%%%%

\section{results\label{sec:results}}

%% put all figures here
\begin{figure*}[t]
\centering
\includegraphics[width=\textwidth]{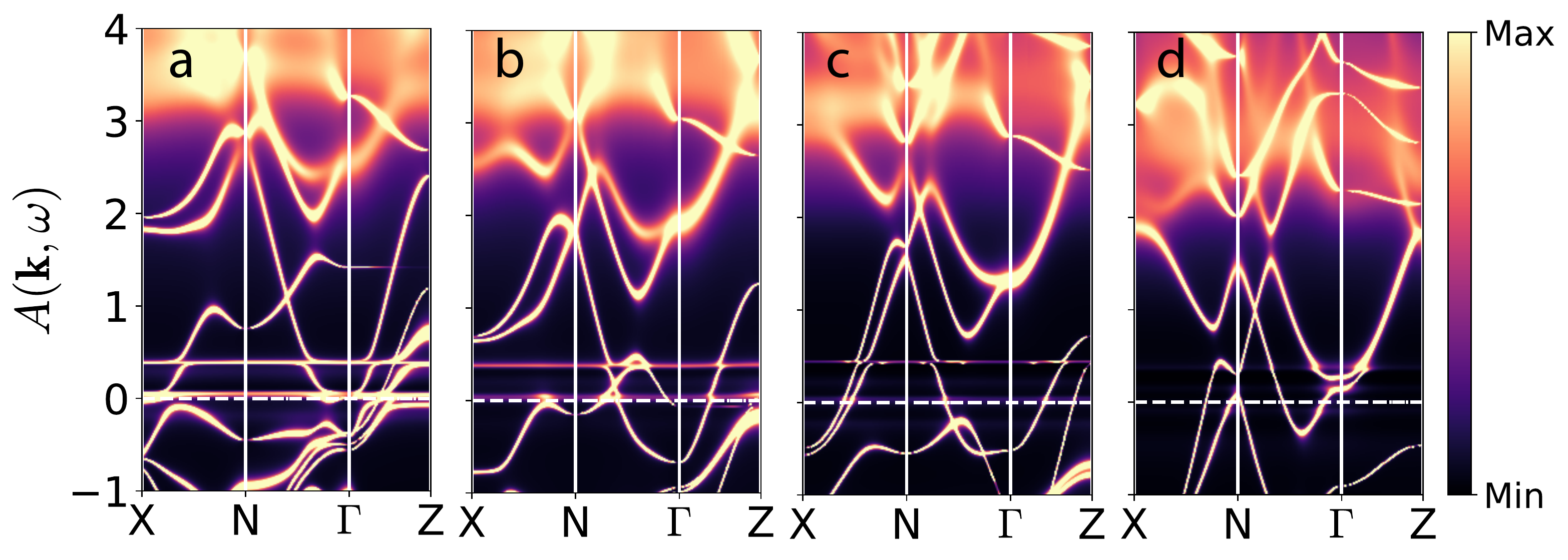}
\caption{(Color online). Momentum-resolved spectral functions $A(\mathbf{k},\omega)$ of Ce$M$$_{2}$Si$_{2}$ obtained by DFT + DMFT calculations. (a) $M$ = Ru. (b) $M$ = Rh. (c) $M$ = Pd. (d) $M$ = Ag. The horizontal dashed lines denote the Fermi level. \label{fig:takw}}
\end{figure*}

\begin{figure*}[t]
\centering
\includegraphics[width=\textwidth]{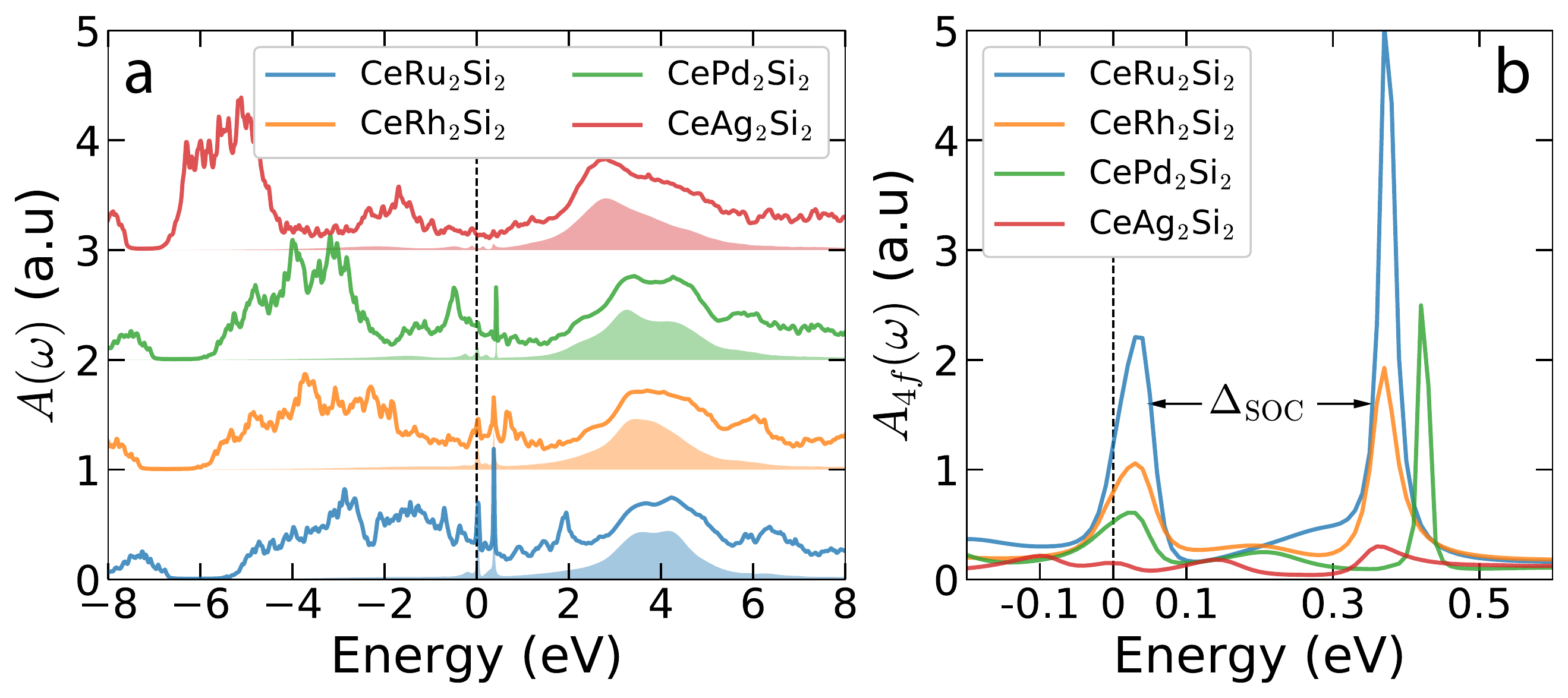}
\caption{(Color online). Electronic density of states of Ce$M$$_{2}$Si$_{2}$ obtained by DFT + DMFT calculations. (a) Total density of states (thick solid lines) and partial 4$f$ density of states (color-filled regions). (b) Partial 4$f$ density of states near the Fermi level. The data presented in this figure are rescaled for a better view. The vertical dashed lines denote the Fermi level. \label{fig:tdos}}
\end{figure*}

\begin{figure}[t]
\centering
\includegraphics[width=\columnwidth]{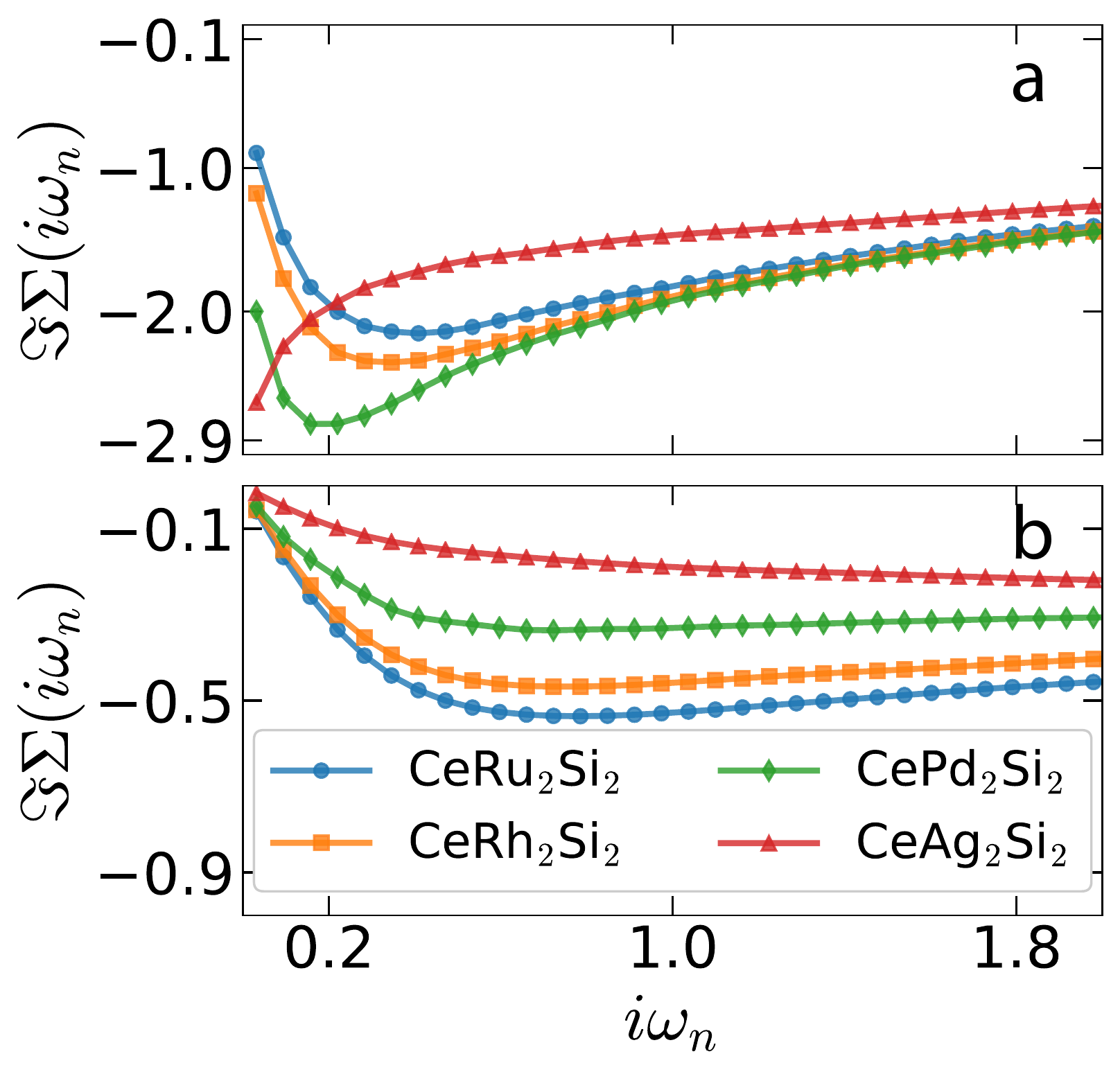}
\caption{(Color online). Imaginary parts of Matsubara self-energy functions of Ce$M$$_{2}$Si$_{2}$ obtained by DFT + DMFT calculations. (a) $4f_{5/2}$ components. (b) $4f_{7/2}$ components. \label{fig:tsigma}}      
\end{figure}

\begin{figure*}[t]
\centering
\includegraphics[width=0.8\textwidth]{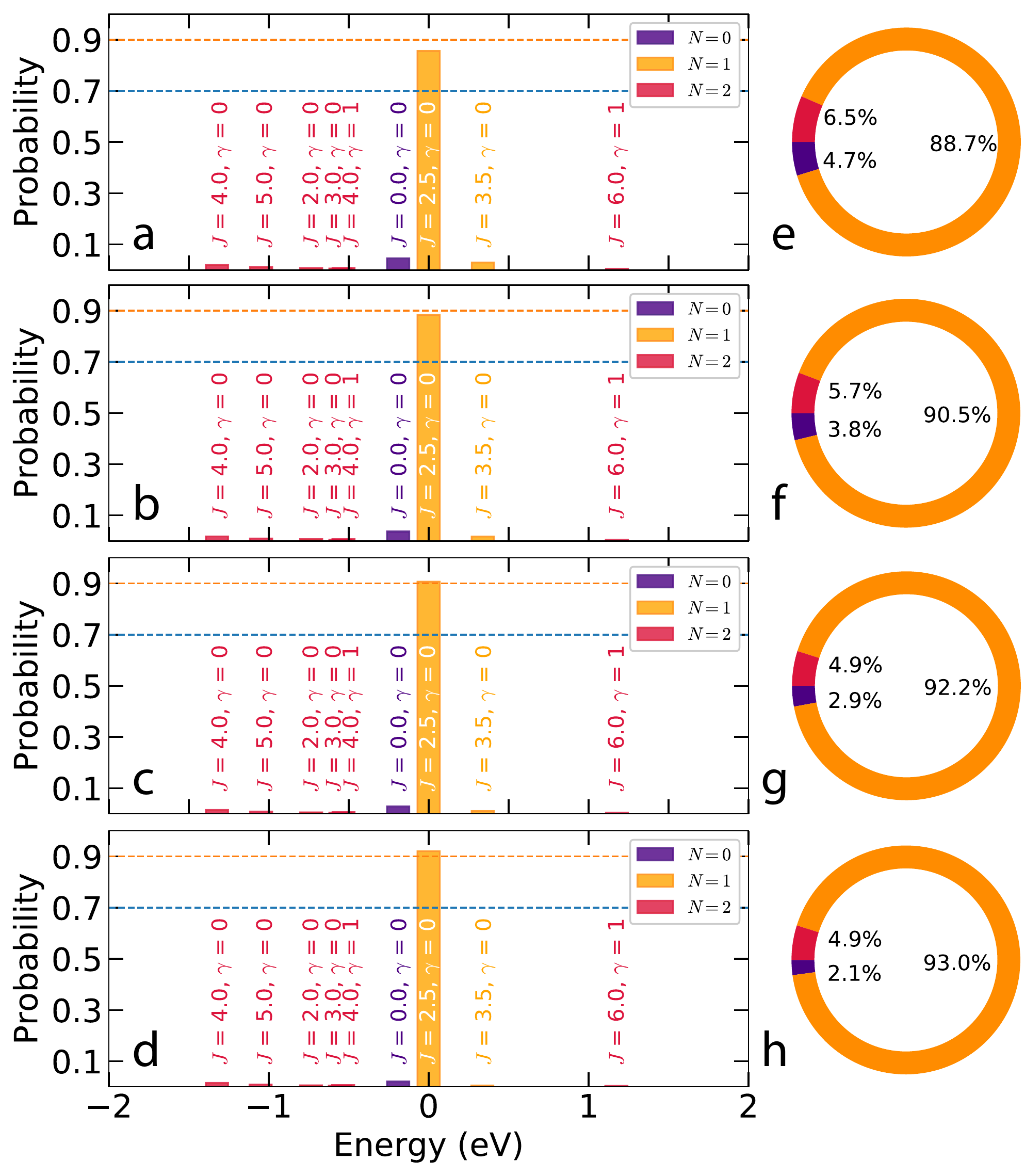}
\caption{(Color online). (a)-(d) Valence state histograms of Ce$M$$_{2}$Si$_{2}$ (where $M =$ Ru, Rh, Pd, and Ag, from top to bottom) by DFT + DMFT calculations. Here we used three good quantum numbers to label the atomic eigenstates. They are $N$ (total occupancy), $J$ (total angular momentum), and $\gamma$ ($\gamma$ stands for the rest of the atomic quantum numbers, such as $J_z$). Note that the contribution from $N = 3$ atomic eigenstates is too trivial to be visualized in these panels. (e)-(h) Probabilities of $4f^{0}$ (violet), $4f^{1}$ (orange), and $4f^{2}$ (red) configurations for Ce$M$$_{2}$Si$_{2}$ (where $M =$ Ru, Rh, Pd, and Ag, from top to bottom) by DFT + DMFT calculations. \label{fig:tprob}}
\end{figure*}

\begin{figure*}[t]
\centering
\includegraphics[width=\textwidth]{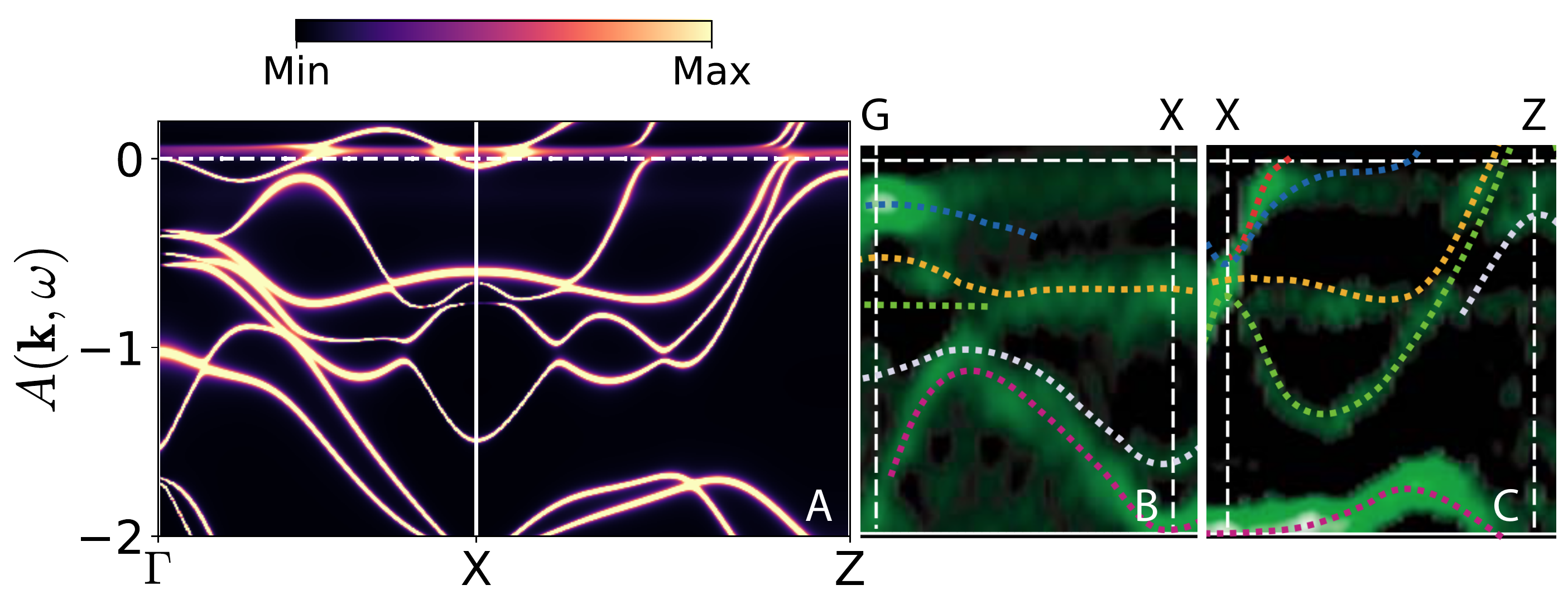}
\caption{(Color online). (a) Momentum-resolved spectral functions $A(\mathbf{k},\omega)$ of CeRu$_{2}$Si$_{2}$ obtained by DFT + DMFT calculations. (b) and (c) ARPES spectra of CeRu$_{2}$Si$_{2}$ measured at 20~K. The colored dashed lines representing each band are guides to the eye. These figures are reproduced from Ref.~[\onlinecite{PhysRevB.77.035118}]. \label{fig:takw_exp}}
\end{figure*}

\begin{figure}[t]
\centering
\includegraphics[width=\columnwidth]{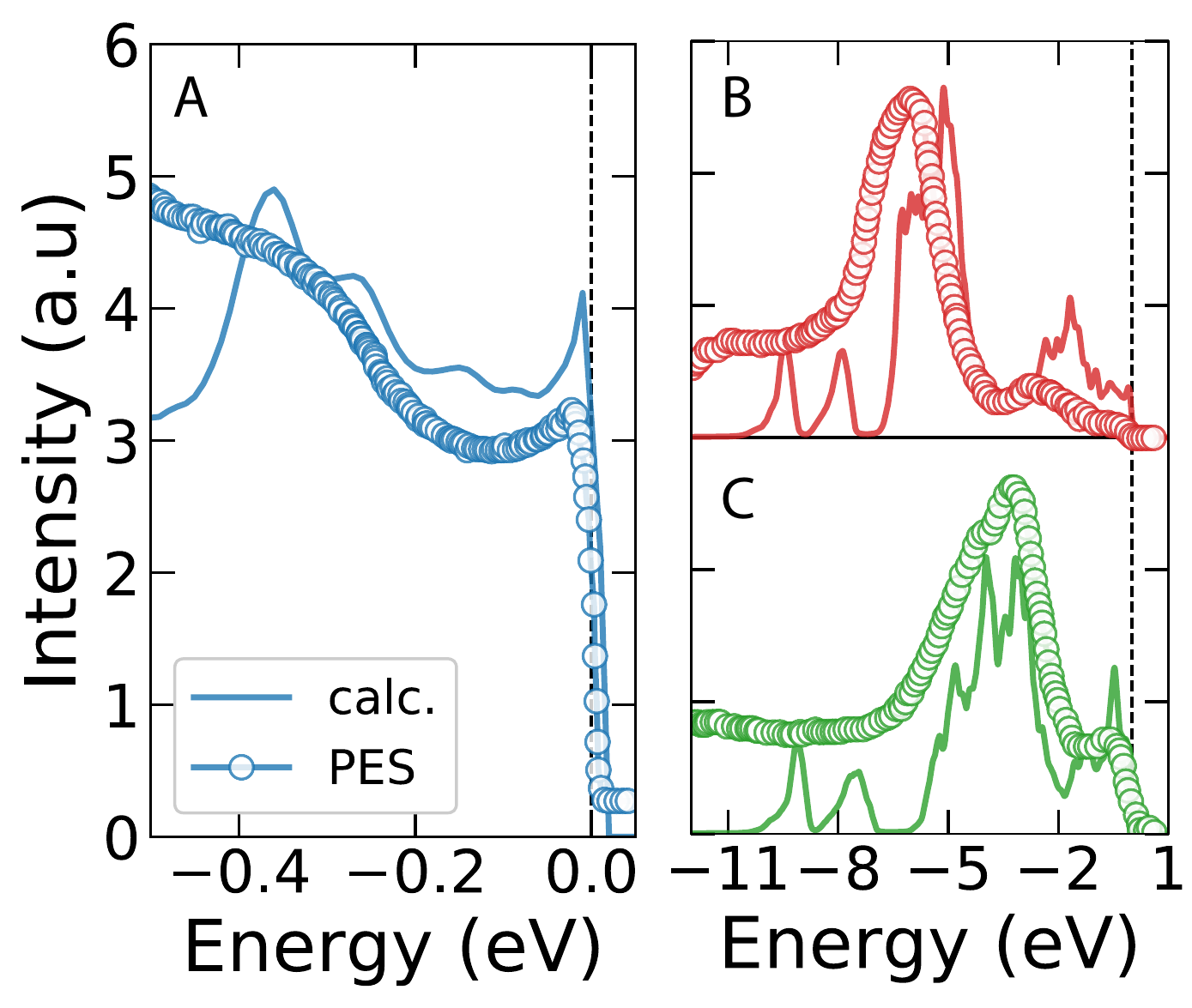}
\caption{(Color online). Electronic density of states of Ce$M$$_{2}$Si$_{2}$. (a) $M =$ Ru. (b) $M =$ Pd. (c) $M =$ Ag. The calculated and experimental data are represented by solid thick lines and empty circles, respectively. The calculated data are multiplied by the Fermi-Dirac distribution function. The experimental data are extracted from Ref.~[\onlinecite{PhysRevB.76.045117}] (for CeRu$_{2}$Si$_{2}$) and [\onlinecite{PhysRevB.27.6052}] (for CePd$_{2}$Si$_{2}$ and CeAg$_{2}$Si$_{2}$). \label{fig:tdos_exp}}
\end{figure}

\subsection{Momentum-resolved spectral functions}

The direct output of self-consistent DFT + DMFT calculations is the Matsubara self-energy functions $\Sigma(i\omega_n)$. They are firstly converted into real-frequency self-energy functions $\Sigma(\omega)$ via analytical continuation procedure~\cite{jarrell}. Then $\Sigma(\omega)$ is used to evaluate the momentum-resolved spectral functions $A(\mathbf{k},\omega)$ and local spectral functions $A(\omega)$. In this subsection, we will pay attention to $A(\mathbf{k},\omega)$ at first.

We tried to calculate the momentum-resolved spectral functions $A(\mathbf{k},\omega)$ of Ce$M_2$Si$_{2}$ along the high-symmetry lines $X - N - \Gamma -Z $ in the irreducible Brillouin zone [see Fig.~\ref{fig:tstruct}(b)]. Figure~\ref{fig:takw} visualizes the calculated results. Surprisingly, though these compounds share similar crystal structures, they display quite different band structures and Fermi surfaces. (i) For $M = $ Ru, Rh, and Pd, the 4$f$ bands dominate when $\omega >$ 3.0~eV. In this energy range, only spread and blurring heat maps are observed. When $\omega <$ 3.0 eV, the $spd$ bands are predominant. They cross the Fermi level and exhibit significant dispersions. For CeAg$_{2}$Si$_{2}$, the 4$f$ bands are closer to the Fermi level. (ii) For CeRu$_{2}$Si$_{2}$, there exist intense and almost flat bands near the Fermi energy, which are associated with the spin-orbit splitting $4f_{5/2}$ and $4f_{7/2}$ bands. The low-lying $4f_{5/2}$ bands locate at the Fermi level, while the high-lying $4f_{7/2}$ bands are at a few hundred meV above the Fermi level. The energy separation between them is approximately 310~meV, which is very close to those observed in Ce metal and some other cerium-based heavy fermion compounds, such as Ce$T$In$_{5}$~\cite{PhysRevLett.108.016402,shim:1615,PhysRevB.81.195107}, CeIn$_{3}$~\cite{PhysRevB.94.075132}, and CeB$_{6}$~\cite{PhysRevB.95.155140}. The prominent $4f_{5/2}$ bands reveal the itinerant behavior of 4$f$ electrons, which is in accord with the paramagnetic ground state and experimentally observed large Fermi surface~\cite{PhysRevB.65.035114,PhysRevLett.101.056401}. When $M$ goes from Ru to Pd, a consequent reduction of the intensities (spectral weights) for these flat bands is observed, connoting the growing localization of 4$f$ electrons. For $M = $ Ag, the flat bands feature near the Fermi level is nearly invisible, which implies that its 4$f$ electrons will approach to the localized limit. (iii) For $M = $ Ru, Rh, and Pd, there exist remarkable $c-f$ hybridizations around the Fermi level. However, for $M = $ Ag, the $c-f$ hybridizations are very weak. (iv) For $M =$ Rh, there is a small Fermi surface pocket (electron type) centered at the $N$ point. While for the other compounds, such pockets are absent. 

\subsection{Density of states}

Next, let us focus on the integrated spectral functions of Ce$M_{2}$Si$_{2}$. Figure~\ref{fig:tdos}(a) shows the total density of states $A(\omega)$ and 4$f$ partial density of states $A_{4f}(\omega)$. Since the spectral weights at the Fermi level are larger than zero, overall the four compounds are metallic. For $M =$ Ru, we can see sharp quasiparticle peak at the Fermi level, which is largely contributed by the $4f_{5/2}$ states. Another more pronounced peak located at $\sim$ 310~meV is mainly associated with the $4f_{7/2}$ states. Note that the ratio of spectral weights of the two peaks is less than 1.0, i.e $I(4f_{7/2})/I(4f_{7/2}) < 1.0$ , which is contrary to those observed in the Ce$T$In$_{5}$ compounds~\cite{PhysRevB.81.195107,shim:1615}. The smooth and broad hump resided from $2$ eV to $6$ eV is mainly assigned to the upper Hubbard bands of cerium's 4$f$ orbitals. On the other hand, the lower Hubbard bands are almost invisible. The density of states of the Ru ion is peaked around binding energy from 1~eV to 5~eV. As for CeRh$_2$Si$_2$ and CePd$_2$Si$_2$, their local spectral functions resemble the one of CeRu$_{2}$Si$_{2}$. The only difference is that the spectral weights of 4$f$ electrons in the vicinity of the Fermi level are transferred to high energy regime. As a result, their quasiparticle peaks become less pronounced. As for CeAg$_{2}$Si$_{2}$, it shows quite different spectral function. At first, the quasiparticle peak almost disappears. Its contribution to the spectral weight at the Fermi level is trivial, indicating the Ag compound is on the localized side of the phase diagram. Second, the upper Hubbard bands are shifted toward the Fermi level. The lower Hubbard bands emerge around -2 eV. They become considerable. Third, the major peaks for the density of states of the Ag ion are moved to -4~eV $\sim$ -7~eV. These differences are consistent with those seen in the momentum-resolved spectral functions [see Fig.~\ref{fig:takw}]. 

Figure~\ref{fig:tdos}(b) zooms in the low-energy part of the 4$f$ partial density of states. We see that CeRu$_{2}$Si$_{2}$ has the highest quasiparticle peak, CeRh$_{2}$Si$_{2}$ follows, and CePd$_{2}$Si$_{2}$ has smaller quasiparticle peak. However, CeAg$_{2}$Si$_{2}$ has no quasiparticle peak left. Only a broad background of the $4f$ spectral weight is seen around the Fermi level. These results suggest that the Ru compound is the most itinerant. Rh and Pd compounds are very similar to Ru compound, but less itinerant. On the other hand, the Ag compound is localized. In Fig.~\ref{fig:tdos}(b), we also find that the peak attributed to the 4$f_{7/2}$ state of CePd$_{2}$Si$_{2}$ is shifted to higher energy, resulting in larger $\Delta_{\text{SOC}}$ (it is equal to the energy level difference between the $4f_{5/2}$ and $4f_{7/2}$ states) than those of the other cerium-based 122 compounds. This abnormal feature is likely attributed to the crystal structure of CePd$_{2}$Si$_{2}$. Actually, it has the largest crystal volume $V$, the smallest $c/a$ ratio, and the longest Ce-Ce distance among the Ru, Rh, and Pd compounds (please refer to Table~\ref{tab:param}). 

\subsection{Self-energy functions}

Figure~\ref{fig:tsigma} shows the Matsubara self-energy functions for 4$f$ states of Ce$M_{2}$Si$_{2}$. In general, we can use the following equation to fit the imaginary part of low-frequency Matsubara self-energy function:
\begin{equation}
-\Im \Sigma (i\omega_n) = A (i\omega_n)^{\alpha} + \gamma. 
\end{equation}
Here, $A$ is a fitting parameter. The exponent parameter $\alpha$ can be used to examine whether the Landau Fermi-liquid theory is fulfilled. According to self-energy data presented in Fig.~\ref{fig:tsigma}, we find that the extracted $\alpha$ parameters are less than 1.0. It manifests the behaviors of 4$f$ electrons in these compounds deviate from the description of the Landau Fermi-liquid theory. The $\gamma$ parameter denotes the low-energy scattering rate of 4$f$ electrons. It is equivalent to $\Im \Sigma(i\omega_n \to 0)$. For the $4f_{7/2}$ states, $\gamma$ approaches zero. While for the $4f_{5/2}$ states, $\gamma$ is much larger than zero. We can concluded that the systems resemble the non-Fermi-liquid state. Furthermore, their quasiparticle weights $Z$ and the electron effective masses $m^{\star}$ should show very strong orbital dependence.

Finally, we notice that for $M = $ Ru, Rh, and Pd, the self-energy functions for both the $4f_{5/2}$ and $4f_{7/2}$ components exhibit metallic features. However, for CeAg$_{2}$Si$_{2}$, the situation is a bit different. Its $4f_{7/2}$ component shows metallic behavior ($Z \approx 0.65$, $m^{\star} \approx 1.55 m_{e} $), while its $4f_{5/2}$ component is insulating ($Z \approx 0.012$, $m^{\star} \approx 84.90 m_{e}$). Clearly, the $4f$ electrons in the $4f_{5/2}$ state are more correlated. We can regard this scenario as an orbital selective 4$f$ insulating state, which is an analogy to the orbital selective Mott phase identified in transition metal compounds, such as Ca$_{2-x}$Sr$_{x}$RuO$_{4}$~\cite{Anisimov2002}. 

\subsection{Valence state fluctuations}

Now let us concentrate on 4$f$ valence state fluctuations and electronic configurations in the four compounds. The CT-HYB quantum impurity solver is capable of computing the valence state histogram (or equivalently atomic eigenstate probability) $p_{\Gamma}$ for 4$f$ electrons, which presents the probability to find out a 4$f$ valence electron in a given atomic eigenstate $|\psi_\Gamma\rangle$ (labeled by good quantum numbers $N$ and $J$ as mentioned in Sec.~\ref{sec:method})~\cite{PhysRevB.75.155113}. Fig.~\ref{fig:tprob}(a-d) illustrate the calculated 4$f$ valence state histograms for Ce$M_{2}$Si$_{2}$. It is easy to notice that the atomic eigenstate $|N = 1, J = 2.5, \gamma = 0\rangle$ is overwhelmingly dominant, followed by the two atomic eigenstates $|N = 0, J = 0.0, \gamma = 0\rangle$ and $|N = 1, J = 3.5, \gamma = 0\rangle$. The probabilities for the remaining atomic eigenstates are negligible. For example, in CeRu$_2$Si$_2$, the probabilities for the three atomic eigenstates are approximately 85.5\%, 4.5\%, and 2.9\%, respectively. As the transition metal ion $M$ varies from Ru to Ag, the probability for the atomic eigenstate $|N = 1, J = 2.5, \gamma = 0 \rangle$ grows up slightly. Accordingly, the probabilities for the atomic eigenstates $|N = 0, J = 0.0, \gamma = 0\rangle$ and $|N = 1, J = 3.5, \gamma = 0\rangle$ are reduced. It is suggested that the redistribution of atomic eigenstates probabilities strongly depends on the atomic number of the transition metal ion $M$. The 4$f$ valence state electrons favor to stay at the ground state $|N = 1, J = 2.5, \gamma = 0\rangle$ more and more.

Since the atomic eigenstates probabilities $p_{\Gamma}$ have been calculated, we can sum up them with respect to $N$ to get the distribution of 4$f$ electronic configurations~\cite{PhysRevB.95.155140,PhysRevB.94.075132,PhysRevB.75.155113}. It will provide some useful information about the 4$f$ valence state fluctuations of the system. Fig.~\ref{fig:tprob}(e-h) shows the distribution of 4$f$ electronic configurations of Ce$M_{2}$Si$_{2}$. Apparently, the $4f^{1}$ configuration always dominates ($\sim 90\%$). The $4f^{2}$ and $4f^{0}$ configurations are less important. They account for $< 7\%$ and $< 5\%$, respectively. The proportion for the $4f^{3}$ configuration is trivial and can be ignored surely. Similar data have been reported for some other cerium-based heavy fermion compounds~\cite{PhysRevB.81.195107, PhysRevB.94.075132,PhysRevB.95.155140}. We find that the proportion of the $4f^{1}$ configuration slightly rises, while those of the $4f^{2}$ and $4f^{0}$ configurations monotonically decline in connection with the atomic number of transition metal ion $M$. It means that the $4f$ valence state fluctuation becomes the most remarkable in CeRu$_{2}$Si$_{2}$. When $M$ goes from Ru to Ag, the 4$f$ valence state fluctuation will be suppressed gradually. 

\subsection{Compared with experimental results}

The experimental results concerning with the electronic structures of the four cerium-based 122 compounds, except CeRu$_{2}$Si$_{2}$, are very limited in the literatures. As a consequence, most of the calculated results presented above can be considered as critical predictions. In this subsection, we would like to compare our results with the available experimental results in order to demonstrate the usefulness and reliability of our methods. 

In Fig.~\ref{fig:takw_exp}, the momentum-resolved spectral function of CeRu$_{2}$Si$_{2}$ is compared with the band structures measured by ARPES~\cite{PhysRevB.77.035118}. The main features observed in the ARPES spectra are successfully reproduced by our DFT + DMFT calculations. In Fig.~\ref{fig:tdos_exp}, the calculated density of states are compared with the experimental photoemission spectra~\cite{PhysRevB.76.045117,PhysRevB.27.6052}. They are roughly in accordance with each other. Let's take the case of CeRu$_{2}$Si$_{2}$ as an example. The comparison is shown in Fig.~\ref{fig:tdos_exp}(a). Both theoretical and experimental spectra show sharp quasiparticle peaks in the vicinity of the Fermi level. As for the 4$f$ electronic configurations, recent linear polarized soft X-ray absorption spectroscopy experiment at $T =$ 20~K suggested that the proportions of $4f^{0}$ configurations for CeRu$_{2}$Si$_{2}$, CeRh$_{2}$Si$_{2}$, and CePd$_{2}$Si$_{2}$ are 6\%, 2.4\%, and 1.4\%, respectively~\cite{PhysRevB.85.035117}. On the other hand, the correspondingly theoretical values are 4.7\%, 3.8\%, and 2.9\%, respectively. Obviously, even though some small discrepancies exist between the theoretical and experimental results, the methods and calculated parameters we used are reliable. Our calculated results are reasonable on the whole.   
  
\subsection{Discussion}

\begin{figure}[t]
\centering
\includegraphics[width=\columnwidth]{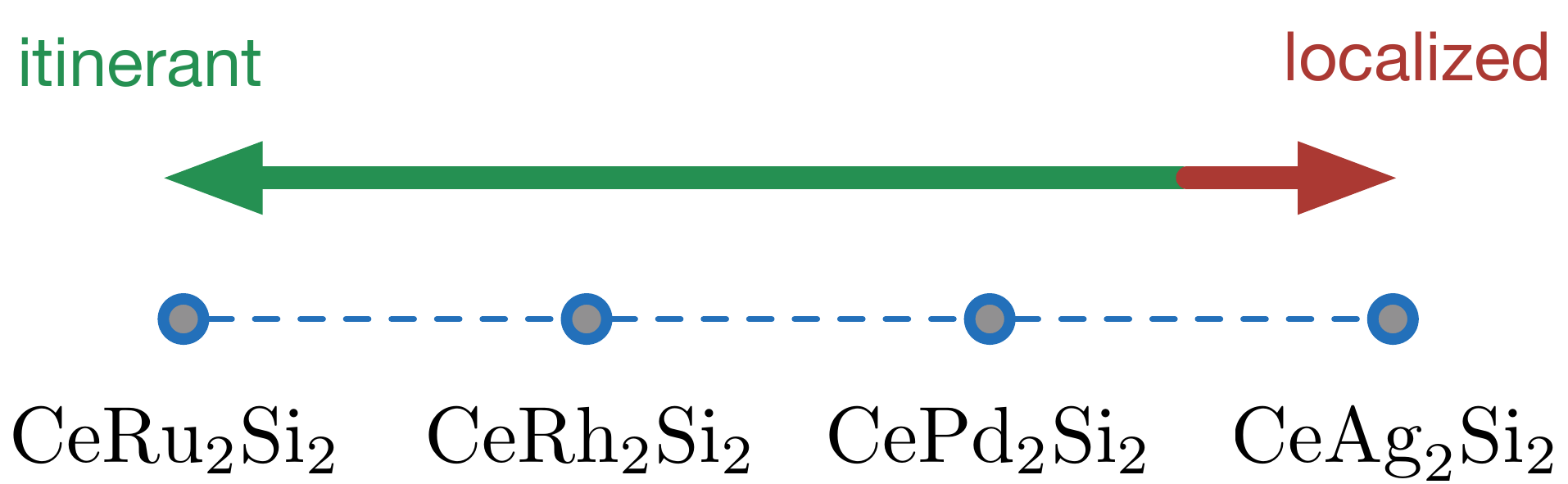}
\caption{(Color online). The sketch of the itinerancy/localization of the four cerium-based 122 compounds. In our view, the 4$f$ itinerant-localized crossover lies between the Pd compound and the Ag compound. \label{fig:tcrossover}}
\end{figure}

\begin{table}[t]
\caption{Crystal structure parameters of Ce$M_2$Si$_{2}$, where $M=$ Ru, Rh, Pd, and Ag~\cite{STEEMAN1988103}. \label{tab:param}}
\begin{ruledtabular}
\begin{tabular}{ccccc}
$M$ & $V$ (${\AA}^3$) & $c/a$ & $d_{\text{Ce-Ce}}$ ({\AA})& $d_{\text{Ce-M}}$ ({\AA}) \\
\hline 
Ru & 192.93 & 2.34 & 4.185 & 3.22 \\
Rh & 190.97 & 2.49 & 4.086 & 3.26 \\
Pd & 198.34 & 2.34 & 4.221 & 3.25 \\
Ag & 214.31 & 2.51 & 4.233 & 3.40 \\
\end{tabular}
\end{ruledtabular}
\end{table}

Based on the calculated results, we can make a preliminary conclusion about the itinerancy or localization of 4$f$ electrons for the four cerium-based 122 compounds (see Fig.~\ref{fig:tcrossover}). We find that the 4$f$ electrons become increasingly localized from $M = $ Ru, to Rh, Pd, and Ag. They are maximum itinerant for $M =$ Ru, and maximum localized for $M = $ Ag. Furthermore, these exists a 4$f$ itinerant-localized crossover between $M =$ Pd and $M =$ Ag. This trend coincide with the increasing atomic number of transition metal ion $M$. 

Next, we would like to seek the underlying mechanism and driving force for the 4$f$ itinerant-localized crossover. Firstly, the four compounds share similar ThCr$_{2}$Si$_{2}$-type crystal structures as stated before (see Fig.~\ref{fig:tstruct}), but with different structural parameters (see Tab.~\ref{tab:param}). We find that none of these structural parameters, including crystal volume, $c/a$ ratio, bond distances between Ce and $M$ atoms, can explain the trend of itinerant to localized crossover in these compounds. Thus, the structure itself is not the driving force of the crossover. Second, the four compounds exhibit considerable 4$f$ valence state fluctuations. The 4$f$ itinerant-localized crossover is accompanied with the change of valence state fluctuation. However, the change is too small to drive an electronic transition. Finally, the cerium-based 122 materials are very sensitive to the substitution of the transition metal ion layer. Besides, the Ru, Rh, Pd, and Ag ions are not isovalent. Thus, we believe that the chemical substitution (or chemical pressure) is indeed the driving force of the itinerant-localized crossover. 

The electronic structures of CeRu$_{2}$Si$_{2}$ have been studied by DFT calculations~\cite{PhysRevB.69.125116,PhysRevB.71.184420}. In the previous calculations, the 4$f$ electrons of cerium are assumed to be fully itinerant. Since the 4$f$ electrons in CeRu$_{2}$Si$_{2}$ are mostly itinerant, this treatment is reasonable. However, traditional DFT method can not be used to study the dual nature of 4$f$ electrons. On the contrary, the DFT + DMFT method provides a reliable tool to study the electronic structures of cerium-based heavy fermion and intermediate valence materials, regardless of the itinerancy or localization of 4$f$ electrons.

%% summary
%%%%%%%%%%%%%%%%%%%%%%%%%%%%%%%%%%%%%%%%%%%%%%%%%%%%%%%%%%%%%%%%%%%%%%%%%%

\section{concluding remarks\label{sec:summary}}

In the present work, we performed \emph{ab initio} many-body calculations to study the electronic structures of four cerium-based 122 compounds, Ce$M_2$Si$_2$, where $M = $ Ru, Rh, Pd, and Ag. We obtained the momentum-resolved spectral functions $A(\mathbf{k},\omega)$, the total and $4f$ partial density of states, Matsubara self-energy functions, and $4f$ valence state fluctuations. We find that from $M = $ Ru, to Rh, Pd, and Ag, the 4$f$ electrons become more and more localized. The compounds with $M =$ Pd and $M = $ Ag stand on the itinerant and localized sides in the phase diagram, respectively. This itinerant-localized crossover is driven by the chemical pressure, and accompanied by change of 4$f$ valence state fluctuation. Of the most interesting is that we identify an orbital selective 4$f$ insulating state in CeAg$_{2}$Si$_{2}$, where the $4f_{5/2}$ states are metallic while the $4f_{7/2}$ states keep insulating. Our results are in closely consistent with the available experiments. Most of the calculated results even serve as useful predictions. 

We would like to point out that the itinerant-localized crossover and 4$f$ valence state fluctuation are common in many rare-earth heavy fermion systems, which has been a longstanding issue and yet to be answered. The study on the electronic structures and valence state fluctuations of typical Ce$M_2$Si$_{2}$ compounds sheds light on the subject, which needs further experimental and theoretical confirmations.

%% funds
%%%%%%%%%%%%%%%%%%%%%%%%%%%%%%%%%%%%%%%%%%%%%%%%%%%%%%%%%%%%%%%%%%%%%%%%%%

\begin{acknowledgments}
This work was supported by Natural Science Foundation of China (No.~11704347 and No.~11504340), the Foundation of President of China Academy of Engineering Physics (Grant No.~YZ2015012), and the Science Challenge Project of China (Grant No.~TZ2016004). The DFT + DMFT calculations were performed on the Delta cluster (in the Institute of Physics, CAS, China).
\end{acknowledgments}

%% reference
%%%%%%%%%%%%%%%%%%%%%%%%%%%%%%%%%%%%%%%%%%%%%%%%%%%%%%%%%%%%%%%%%%%%%%%%%%

\bibliography{ce122}

\end{document}